\documentclass{article}

\usepackage{graphicx,color,psfrag}
\usepackage{arxiv}

\usepackage[backend=biber,style=apa]{biblatex}
\bibliography{plwh,Books,data}
\AtEveryBibitem{\clearfield{doi}}
\AtEveryBibitem{\clearfield{isbn}}
\AtEveryBibitem{\clearfield{issn}}

\DeclareNameAlias{sortname}{family-given}
\renewbibmacro{in:}{}
\DeclareFieldFormat[article, inbook, incollection, inproceedings, misc, thesis, unpublished]{title}{#1}
\DeclareFieldFormat[article]{volume}{{#1}} 
\DeclareFieldFormat[article, inbook, incollection, inproceedings, misc, thesis, unpublished]{number}{\mkbibparens{#1}} 
\renewbibmacro*{volume+number+eid}{
  \printfield{volume}
  \printfield{number}
}

\title{The Best Thresholds for Rapid Identification of Episodic and Chronic Homeless Shelter Use}
\author{Geoffrey Messier  \\
  University of Calgary  \\
  2500 University Dr.~NW, Calgary, AB, Canada, T2N 1N4\\
  gmessier@ucalgary.ca
  \and
  Leslie Tutty  \\
  University of Calgary  \\
  2500 University Dr.~NW, Calgary, AB, Canada, T2N 1N4\\
  tutty@ucalgary.ca
  \and
  Caleb John\\
  University of Calgary  \\
  2500 University Dr.~NW, Calgary, AB, Canada, T2N 1N4\\
  ctjohn@ucalgary.ca
}

\date{}

\newcommand{\bE}{\begin{enumerate}}
\newcommand{\eE}{\end{enumerate}}
\newcommand{\bI}{\begin{itemize}}
\newcommand{\eI}{\end{itemize}}
\newcommand{\I}{\item}

\begin{document}

\maketitle

\begin{abstract}
This paper explores how to best identify clients for housing services based on their homeless shelter access patterns.  We focus on counting the number of shelter stays and episodes of shelter use for a client within a time window.  Thresholds are then applied to these values to determine if that individual is a good candidate for housing support.  Using new housing referral impact metrics, we explore a range of threshold and time window values to determine which combination both maximizes impact and identifies good candidates for housing as soon as possible.  New insights are also provided regarding the characteristics of the ``under-the-radar'' client group who are typically not identified for housing support.
\end{abstract}

\noindent
{\bf Keywords:}  chronic homelessness, episodic homelessness, housing first services, data analysis

\section{Introduction}
\label{sec:intro}

An emergency homeless shelter sees a diversity of clients who access the shelter in different ways and require different services.  Shelter access is often characterized in terms of total stays and number of episodes \parencite{kuhn-r-1998}.   The analysis of shelter stay and episode data has led to the discovery of three different types of shelter access patterns: transitional/temporary, episodic and chronic.  The majority of new clients are transitional and exit shelter after a relatively short time.  Episodic clients access shelter intermittently over a long period and chronic clients access shelter very regularly over a long period.  These three groups have been identified using cluster analysis on shelter access data from a number of North American cities \parencite{kuhn-r-1998, culhane-dp-2007, aubry-t-2013, kneebone-r-2015}.

The original purpose of shelter stay and episode data was to gain a better understanding of the homeless population.  However, it is now also increasingly being used to help guide program delivery.  Housing First services have been a priority for the Canadian federal government's Homelessness Partnering Strategy since 2013 \parencite{gaetz-s-2016}. Housing First is a recovery-oriented approach to quickly and successfully connect individuals and families experiencing homelessness to permanent housing without preconditions and barriers to entry, such as sobriety, treatment or service participation requirements \parencite{goering-p-2014}.  Housing chronically and episodically homeless individuals is one of the priorities of Housing First programming \parencite{gaetz-s-2014, aubry-t-2015}.  A number of definitions for chronic and episodic homelessness have been developed that can applied to shelter access data, used to identify these individuals and then connect them to housing supports \parencite{byrne-t-2015, di-community-report-2019, synder-sg-2008, employment-social-development-2019}.

The purpose of this paper is to examine how best to use stay/episode threshold tests to identify clients for housing services based on their shelter access patterns.  While it is only one of the tools that should be use to triage clients for housing services \parencite{aubry-t-2015}, shelter access data analysis is important since it has the potential to spot clients who are in need help and might otherwise escape notice.  This is particularly true in very busy shelters where staff may not have personal knowledge of all clients.

There are two priorities when identifying an individual for housing supports: selecting the clients in greatest need of housing and finding them as quickly as possible after they first enter shelter.  The primary contribution of this paper is to create new threshold-based tests for chronic and episodic shelter use with these two priorities in mind.  Using 13 years of anonymized client data from a major North American shelter, we will demonstrate that our approach is able to identify clients for housing services almost 200 days sooner than existing definitions for chronic homelessness.  

In the following, Section~\ref{sec:data} describes the dataset used for this study and cluster analysis is used in Section~\ref{sec:cluster} to demonstrate that our data is representative of homeless populations in other major urban centres.  Section~\ref{sec:def} presents the analysis we use to develop our tests which we then compare to existing definitions for chronic homelessness in Section~\ref{sec:results}.  Section~\ref{sec:missing} provides some insight into the individuals who are not identified for housing support and concluding remarks are made in Section~\ref{sec:concl}.

\section{Data Set}
\label{sec:data}

This is a secondary data analysis performed on anonymized shelter stay records collected at the Calgary Drop-In Centre (DI) between July 1, 2007 and January 20, 2020.  The data anonymization and client privacy protocol used during this study was approved by the University of Calgary Conjoint Faculties Research Ethics Board.  The dataset consists of 5,431,521 entries for 34,577 unique client profiles.

Each time a shelter client accesses day sleep or night sleep facilities at the DI, an entry is created in the database with a timestamp and the unique identification number of the client.  For the purposes of this study, a single {\em stay} is defined as a 24 hour period where a client accesses day sleep services, night sleep services or both.  An {\em episode} is a series of stays where the separation between consecutive stay dates is less than 30 days \parencite{byrne-t-2015}.

As discussed in \parencite{kuhn-r-1998}, studies that examine periods of time in shelter suffer from the censoring of client access patterns that begin before the start of the data record (left-censoring bias) and continue after the end of the data record (right-censoring-bias).  Following the approach in \parencite{kuhn-r-1998}, we minimize the effect of left and right censoring bias by only including clients who first appear in the data after July 1, 2009 and prior to January 20, 2018.  A total of 18,398 client profiles meet this inclusion criterion (53.2\% of the original dataset population).

\section{Cluster Analysis}
\label{sec:cluster}

In order for the analysis and conclusions presented in this paper to be generally relevant, it is important to establish that our dataset is representative of homeless populations in other jurisdictions.  To compare our data with existing populations, we perform a cluster analysis in this section to study the nature of our transitional, episodic and chronic populations.

Utilizing the definition for stays and episodes described in Section~\ref{sec:data}, the total number of stays and episodes experienced by each of the $N$ = 18,398 clients included in this study is calculated.  This data is then analyzed using the $k$-means clustering algorithm \parencite{hastie-t-2017, kuhn-r-1998, aubry-t-2013, kneebone-r-2015} where $k = 3$ clusters is assumed.  Z-score variable standardization is used to normalize the stay and episode data to have zero mean and unit variance prior to clustering \parencite{milligan-gw-1988}.  The average stay and episode values for each cluster are given in Table~\ref{tb.clusters}.  Similar to \parencite{aubry-t-2013, kneebone-r-2015}, we validate our clusters by verifying that the differences between the means of our centroids are statistically significant.  Since this is a multi-variable analysis, we utilize Hotelling's t-squared test \parencite{afifi-a-2011}.  All comparisons of the means provided in Table~\ref{tb.clusters} using Hotelling's test yield a p-value of less than 0.001.

\begin{table}[htbp]
  \centering
  \begin{tabular}{c|ccc}
    Group & Average Total Episodes & Average Total Stays & Proportion of Population \\ \hline
    Transitional & 1.8 & 30.3 & 85.2\% (15675/18398) \\ \hline
    Episodic & 9.2 & 167.0 & 11.9\% (2184/18398) \\ \hline
    Chronic & 3.7 & 1273.1 & 2.9\% (539/18398) \\
  \end{tabular}
  \caption{Statistical averages for cluster groups.}
  \label{tb.clusters}
\end{table}

To verify that our dataset is representative, the results in Table~\ref{tb.clusters} can be compared to cluster analysis studies conducted for other North American cities \parencite{kuhn-r-1998, culhane-dp-2007, aubry-t-2013}, including a multi-shelter study conducted in Calgary \parencite{kneebone-r-2015}.  When comparing with the Canadian cities, the proportions of chronic, episodic and transitional clients are similar to within a few percentage points.   The average number of episodes and stays are also similar with exception of the 1273.1 average total stays for chronic clients.  The most similar average number of chronic stays are 927.1 and 769.3 for the multi-shelter Calgary study and Ottawa study, respectively.  Our larger value can be explained by the fact that the DI is the largest shelter in Calgary and, before the advent of Housing First programming, had a number of very regular and long term clients.

When comparing to cities in the United States, our results are most similar to \parencite{kuhn-r-1998}.  The results in \parencite{culhane-dp-2007} have a higher proportion of chronic clients (in some cases up to 20\%) and a higher number of stays in shelter for transitional users (over 100 for Philadelphia and Massachusetts).  We do not think these differences are large enough to invalidate the main conclusions of our work for American cities since the variation between Table~\ref{tb.clusters} and the results in \parencite{culhane-dp-2007} is similar to the variation observed between \parencite{culhane-dp-2007} and \parencite{kuhn-r-1998,  aubry-t-2013, kneebone-r-2015}.

\section{Identifying Chronic Shelter Users}
\label{sec:def}

One way to identify a chronic shelter user is to see if they satisfy one of the many definitions of chronic homelessness \parencite{employment-social-development-2019, synder-sg-2008, di-community-report-2019, byrne-t-2015}.  While these definitions differ in minor ways, they are all variations of applying a threshold test to the number of shelter stays or episodes of shelter use that occur within a time window.  For example, an individual satisfies the Canadian federal definition of homelessness if they have 180 or more shelter stays in one year (a threshold of 180 stays and a time window of 365 days).

When designing a test to identify chronic shelter users for housing support, the test should maximize the impact of each housing intervention.  Section~\ref{ssec:metric} presents a new way to examine historical shelter data to quantify how much time in shelter could be saved when a client is referred to a housing program.  Section~\ref{ssec:opt} then evaluates a large number of time window and stay/episode threshold values to find the combination that maximizes the impact of housing referrals while identifying clients for assistance as quickly as possible.

We will maintain separate definitions for chronic and episodic shelter clients.  While some definitions test separately for chronic and episodic clients \parencite{employment-social-development-2019, synder-sg-2008}, they then lump these clients together and call them all ``chronic''.  We choose to maintain separate tests for episodic and chronic shelter users since they exhibit very different characteristics \parencite{kuhn-r-1998, aubry-t-2013} and could benefit from different types of support.

\subsection{Evaluating the Impact of a Housing Intervention}
\label{ssec:metric}

As noted in Section~\ref{sec:intro}, one priority when delivering housing services is achieving high program efficiency which is defined as ensuring resources are allocated to those who need it most \parencite{shinn-m-2019}.  This could be achieved by using a very conservative test for chronic shelter use that requires a long period in shelter before triggering a housing referral.  However, this approach is at odds with our second priority of quickly providing clients with assistance before the difficult conditions experienced during a long period in shelter have an adverse effect on their mental and physical health.

Finding a test for chronic/episodic shelter use that achieves the best tradeoff between program efficiency and rapid intervention requires the right performance metric.  In the past, the best test for chronic homelessness has been selected as the one identifying the group that consumes the highest percentage of total shelter stays \parencite{byrne-t-2015}.  While this is a good metric for program efficiency, it does not account for how quickly those clients are identified.

To address this point, we propose a metric that evaluates how much a housing referral reduces a client's time in shelter.  We define {\em referral date} as the point in time where a test for chronic or episodic shelter use is satisfied in a client's historic shelter access timeline.  We make the idealized assumption that the client receives a ``perfect'' housing referral on their referral date.  This means the client transitions directly to housing on the same day they are identified by the test and does not return to shelter.

We then calculate the {\em shelter stays saved} and {\em shelter tenure reduction} for that client due to the housing referral.  Consistent with Section~\ref{sec:data}, a {\em stay} is defined as a 24 hour period where the client accesses shelter sleep services one or more times.  The {\em tenure} of a client in shelter is defined as the number of days between the first and last day the client appears in the shelter record.  To determine stays saved by a housing referral, we add up all shelter stays that remain in the client's timeline after the referral date.  To determine the tenure reduction achieved by a housing referral, we count the number of days between the referral date and the last day they appear in their shelter access record.

For example, consider a hypothetical client in the DI dataset that stays in shelter on April 10, 12, 13, 20, 22 and 30.  Assume we apply a test to that access pattern that identifies the client as chronic on April 13.  April 13 would be the client's referral date and we assume they leave shelter and move into a house on that day.  Since there are 3 more stays in the timeline (April 20, 22 and 30), the number of stays saved by the referral is 3.  Since the final time the client actually accesses shelter is April 30, the tenure reduction for the client is 17 days.

While this metric is clearly ideal, it is very effective for evaluating the tradeoff between program efficiency and speed of identification.  For example, a very conservative test that only identifies a client after a long time in shelter will have a low stays saved/tenure reduction score since there will not be much time left in the client's shelter timeline by the time they are identified for a referral.  Conversely, a definition that identifies clients too early will also score low since it will start to identify transitional clients that have a small number of shelter stays in their timeline.

We believe that it is important to evaluate both stays saved and reduction in tenure.  The number of stays saved maps easily to shelter system cost savings since stays are directly linked to the number of beds a shelter must operate.  Calculating reduction in tenure, we argue, is a better measure of social cost.  An individual who interacts with an emergency shelter for an extended period, even if that interaction is very intermittent, is most likely in some level of distress for that entire period and not just the days physically spent in shelter.

\subsection{Finding the Best Test}
\label{ssec:opt}

While existing definitions for chronic homelessness often make intuitive sense, they have not been proven to be ``best'' in any quantifiable way.  In this section, we evaluate several different combinations of stay/episode threshold and time window values to find the combination that achieves the best tradeoff between program efficiency and speed of identification.  

For chronic shelter use, we evaluate the performance of all possible combinations of time windows equal to 30, 90, 180, 365 and 547 days and stay thresholds equal to 50\%, 75\% and 90\% of the time window being applied (a total of 15 threshold/window combinations).  For episodic shelter use, we evaluate all combinations of time windows equal to 30, 90, 180, 365 and 547 days and episode thresholds equal to 2, 3, 4 and 5 episodes (a total of 20 threshold/window combinations).  We choose to evaluate the performance of the chronic test in terms of average stays saved per housing referral since chronic clients make the heaviest use of system resources.  Episodic test performance is evaluated in terms average tenure reduction per housing referral since that emphasizes the extended period of time episodic clients may be in distress.

Tables~\ref{tb.optChr} and \ref{tb.optEpi} show the top ten performing window/threshold combinations for the chronic and episodic tests, respectively.  The tables also include the number of clients identified, percentage of total clients identified and the median time to identification.  

\begin{table}[htbp]
  \centering
  \begin{tabular}{cccc}
   Window (days)/ &  & Avg. Stays Saved & Median ID Time \\
   Threshold (stays) & N & per Referral & (days) \\ \hline
547/492 & 415 (2.3\%) & 721.4 & 538.0 \\
365/328 & 594 (3.2\%) & 687.9 & 376.5 \\
547/410 & 661 (3.6\%) & 656.6 & 496.0 \\
365/273 & 904 (4.9\%) & 619.5 & 350.5 \\
180/162 & 1075 (5.8\%) & 588.4 & 222.0 \\
547/273 & 1138 (6.2\%) & 555.7 & 421.5 \\
180/135 & 1583 (8.6\%) & 504.4 & 188.0 \\
365/182 & 1536 (8.3\%) & 499.7 & 287.0 \\
90/81 & 1815 (9.9\%) & 471.2 & 137.0 \\
180/90 & 2370 (12.9\%) & 416.6 & 142.0 \\
   \hline
  \end{tabular}
  \caption{Window/threshold performance for chronic detection.}
  \label{tb.optChr}
\end{table}

\begin{table}[htbp]
  \centering
  \begin{tabular}{cccc}
   Window (days)/ &  & Avg. Tenure Reduction  &  Median ID Time\\
   Threshold (episodes) & N & per Referral (days) & (days) \\ \hline
180/4 & 310 (1.7\%) & 932.9 & 731.5 \\
365/5 & 618 (3.4\%) & 906.6 & 748.5 \\
365/3 & 3476 (18.9\%) & 872.8 & 313.0 \\
547/5 & 1246 (6.8\%) & 868.1 & 666.5 \\
180/3 & 1903 (10.3\%) & 864.7 & 460.0 \\
547/4 & 2331 (12.7\%) & 858.1 & 485.0 \\
365/4 & 1581 (8.6\%) & 857.0 & 569.0 \\
547/3 & 4201 (22.8\%) & 854.7 & 367.0 \\
90/2 & 4277 (23.2\%) & 829.7 & 181.0 \\
90/3 & 275 (1.5\%) & 829.3 & 663.0 \\
\hline
  \end{tabular}
  \caption{Window/threshold performance for episodic detection.}
  \label{tb.optEpi}
\end{table}

When deciding which window/threshold combinations to select from Tables~\ref{tb.optChr} and \ref{tb.optEpi}, we are mindful of the priority of identifying clients for assistance as quickly as possible.  In Table~\ref{tb.optChr}, a window of 547 days and a threshold of 492 stays has the best performance in terms of stays saved per housing referral.  This test maximizes stays saved since it selects only the 2.3\% of clients who are heavy, long term shelter users.  However, this test is unacceptable because it takes a median time of 538~days (approx. 1.5 years) to identify a client for referral.

If we examine Table~\ref{tb.optChr} at a higher level, we can see that even the worst performing test still saves an impressive 416.6 shelter stays per referral.  Since all the tests are very high impact, we are free to choose the test that also offers rapid identification of chronic shelter users.  A time window of 90 days and a stay threshold of 81 offers a median time to identification of only 137 days (less than 5 months).  This test ranks 9th but still selects a cohort of clients for referrals that save an average of 471.2 stays each time a client is placed in housing.

When examining the episodic results in Table~\ref{tb.optEpi}, the tenure reduction values are very strong for all of the window/threshold combinations.  Therefore, we select the 90 day window and 2 episode threshold combination as achieving a good tradeoff between impact and speed of detection.  This test ranks 9th but still reduces a client's tenure of shelter interaction by an average of 829.7 days (2.4 years) when that client is referred to housing.

To summarize, we would like to propose the RAPid IDentification (RAPID) test for chronic shelter use (RAPID-Chronic) as identifying any client with 81 or more stays in shelter in a 90 day period.  The RAPID test for episodic shelter use (RAPID-Episodic) identifies any client with 2 or more episodes of shelter access in a 90 day period.

\section{Evaluation and Results}
\label{sec:results}

In this section, we compare the RAPID tests to existing definitions for chronic homelessness.  We select the following definitions as being generally representative of the homelessness serving sector and relevant to our local, provincial and national contexts:

\bI
\I {\bf Government of Canada (GoC):} ``individuals who are currently experiencing homelessness and who meet at least 1 of the following criteria: they have a total of at least 6 months (180 days) of homelessness over the past year or they have recurrent experiences of homelessness over the past 3 years, with a cumulative duration of at least 18 months (546 days)'' \parencite{employment-social-development-2019}

\I {\bf Government of Alberta (GoA):} ``A person or family is considered chronically homeless if they have either been continuously homeless for a year or more, or have had at least four episodes of homelessness in the past three years.'' \parencite{synder-sg-2008}
\eI

In the following, Section~\ref{ssec:clientId} presents the demographics of the individuals selected for housing by the RAPID, GoC and GoA definitions. Section~\ref{ssec:compare} compares how each definition performs in terms of stays saved, tenure reduction and time to identification.

\subsection{Clients Identified for Support}
\label{ssec:clientId}

The client group statistics presented in this section include total shelter stays, total episodes of shelter access, tenure of shelter interactions and shelter usage percentage (total stays divided by tenure).  Each table also indicates the total coverage of a definition (number of clients identified).  The statistics of the clients flagged by the GoC and GoA are given in Tables~\ref{tb.gocFlg}, and \ref{tb.goaFlg}, respectively.  The RAPID-Chronic and RAPID-Episodic tests are shown in Tables~\ref{tb.prgChrFlg} and \ref{tb.prgEpiFlg}, respectively.

The results in Table~\ref{tb.gocFlg} indicate that the GoC definition selects primarily chronic clients.  It is interesting to note that the intention of the GoC definition was also to lump in episodic shelter users by including a second stay count threshold for a three year period. However, this second test still selects primarily chronic clients.  In contrast, the GoA test lumps together clients that satisfy either a stay count test or episode test.  This does effectively group episodic and chronic users together.  As a result, Table~\ref{tb.goaFlg} shows a higher percentage of clients selected, lower values for total stays, higher episode counts and lower shelter usage percentages.

The RAPID-Chronic results in Table~\ref{tb.prgChrFlg} are most similar to the GoC definition in terms of shelter usage percentage, tenure and episode counts.  The values in Table~\ref{tb.prgEpiFlg} for RAPID-Episodic reflect the primarily episodic clients selected by this test with low stay counts but high tenure and episode counts.

\begin{table}[htbp]
  \centering
  \begin{tabular}{l|ccc}
    & Mean & Median & 90th Percentile \\ \hline
    Total Stays & 702.7 & 522.0 & 1430.0 \\
    Total Episodes & 4.4 & 3.0 & 9.0 \\
    Tenure (days) & 1564.1 & 1439.0 & 2940.0 \\
    Usage Percentage & 53.0\% & 48.6\% & 95.8\% \\
    \hline
    \multicolumn{4}{c}{Coverage: 1549/18398 (8.4\%)} \\
    \hline
  \end{tabular}
  \caption{Government of Canada definition.}
  \label{tb.gocFlg}
\end{table}

\begin{table}[htbp]
  \centering
  \begin{tabular}{l|ccc}
    & Mean & Median & 90th Percentile \\ \hline
    Total Stays & 438.6 & 216.0 & 1128.0 \\
    Total Episodes & 7.4 & 7.0 & 14.0 \\
    Tenure (days) & 1672.2 & 1592.0 & 2974.0 \\
    Usage Percentage & 28.3\% & 14.1\% & 84.1\% \\
    \hline
    \multicolumn{4}{c}{Coverage: 2443/18398 (13.3\%)}\\
    \hline
  \end{tabular}
  \caption{Government of Alberta definition.}
  \label{tb.goaFlg}
\end{table}

\begin{table}[htbp]
  \centering
  \begin{tabular}{l|ccc}
    & Mean & Median & 90th Percentile \\ \hline
    Total Stays & 601.0 & 411.0 & 1329.0 \\
    Total Episodes & 4.0 & 3.0 & 9.0 \\
    Tenure (days) & 1399.9 & 1241.0 & 2837.0 \\
    Usage Percentage & 53.1\% & 49.1\% & 96.2\% \\
    \hline
    \multicolumn{4}{c}{Coverage: 1815/18398 (9.9\%)}\\
    \hline
  \end{tabular}
  \caption{RAPID-Chronic test.}
  \label{tb.prgChrFlg}
\end{table}

\begin{table}[htbp]
  \centering
  \begin{tabular}{l|ccc}
    & Mean & Median & 90th Percentile \\ \hline
    Total Stays & 136.9 & 39.0 & 376.0 \\
    Total Episodes & 6.3 & 5.0 & 12.0 \\
    Tenure (days) & 1308.4 & 1149.0 & 2763.0 \\
    Usage Percentage & 11.2\% & 5.2\% & 30.4\% \\
    \hline
    \multicolumn{4}{c}{Coverage: 4277/18398 (23.2\%)}\\
    \hline
  \end{tabular}
  \caption{RAPID-Episodic test.}
  \label{tb.prgEpiFlg}
\end{table}

\subsection{Definition Performance Comparison}
\label{ssec:compare}

Table~\ref{tb.prf} compares the different definitions in terms of number of clients identified, stays saved per referral, tenure reduction per referral, average time to identification and median time to identification.  The RAPID results are generated by applying both the RAPID-Chronic and RAPID-Episodic tests to each client in the DI dataset and then classifying them according to whichever test is satisfied earliest in the client's timeline.  As a result, the total clients identified by RAPID in Table~\ref{tb.prf} is less than the sum of clients in Tables~\ref{tb.prgChrFlg} and \ref{tb.prgEpiFlg} since some clients satisfy both RAPID-Chronic and RAPID-Episodic.

\begin{table}[htbp]
  \centering
  \begin{tabular}{l|ccccc}
    & Clients    & Stays Saved  &  Tenure Reduction& Mean Time & Median Time  \\
    & Identified & per Referral & per Referral (days) & to ID (days) & to ID (days)\\ \hline
    Government of Alberta & 2443 (13.3\%) & 264.8 & 914.2 & 609.8 & 365.0 \\ 
    Government of Canada & 1549 (8.4\%) & 498.4 & 1007.9 & 555.2 & 285.0 \\
    RAPID & 5507 (30.0\%) & 194.8 & 874.4 & 402.2 & 98.0 \\ 
    \hline
  \end{tabular}
  \caption{Definition performance comparison.}
  \label{tb.prf}
\end{table}

With a median time to identification of only 98 days or just over 3 months, we think that the RAPID tests are the clear choice for housing program delivery.  The next quickest definition requires that a client remain in shelter for 285 days before being identified for support.  While the RAPID tests result in the lowest number of stays saved and lowest tenure reduction in Table~\ref{tb.prf}, they still have an impressive impact with an average number of stays saved per referral of 194.8 and an average tenure reduction per referral of 874.4 days.  Clearly, RAPID does a good job of avoiding short term transitional shelter users and finding high impact clients to connect to housing services.

Since RAPID identifies the highest proportion of clients, one possible concern is that the RAPID tests will flag too many clients and overload housing programs with potential candidates.  When applied to the DI dataset, RAPID identifies an average of just over 44 individual for housing services per month.  The current housing programs at the Calgary Drop-In Centre have successfully been able to house approximately this many clients \parencite{di-community-report-2019}.  Therefore, we believe that the number of clients identified by RAPID should fall within the capacity of most housing programs.  It should also be stressed that clients identified by RAPID are not necessarily automatically placed in a home.  As discussed in Section~\ref{sec:intro}, these tests should be considered only one component of a housing program that includes discussions with each client to determine how best to support their exit from the shelter system.

Finally, we point out that Table~\ref{tb.prf} is a dramatic illustration of the potential impact of Housing First services.  Regardless of which definition is used, housing referrals have the potential to save many hundreds of shelter stays and, in some cases, over 1,000 days of tenure in shelter {\em per referral}.

\section{Who are we missing?}
\label{sec:missing}

While it is important to analyze who is being identified for housing support, it is equally important to evaluate who is being missed.  The statistics for the transitional clients not flagged by RAPID are shown in Table~\ref{tb.pdTmp}.  Consistent with our understanding of transitional clients, as discussed in Section~\ref{sec:cluster}, Table~\ref{tb.pdTmp} shows a client population with a small number of total stays and episodes.  The shelter usage percentages are high, primarily due to the large number of clients with a single stay which corresponds to a 100\% shelter use percentage (1 stay divided by 1 day).

The most significant item to highlight in Table~\ref{tb.pdTmp} is the 90th percentile tenure value.  This group of clients, corresponding to 1289 individuals or 7.0\% of the total client population, has a period of shelter interaction of 1450 or more days.  As acknowledged in Section~\ref{ssec:metric}, these are not days of continual interaction and likely contain large gaps where the client is absent from the shelter system.

To explore this further, the clients in the 90th percentile of Table~\ref{tb.pdTmp} are analyzed as a separate cohort.  Within this group, the average number of days between episodes of shelter access is 635.8.  This creates a picture of a population who interact with an emergency shelter over a period of several years with gaps in that interaction of longer than a year.  These clients do not use the shelter system intensively enough to be flagged as chronic or episodic but obviously are existing for very long periods in precarious circumstances.  At 7.0\% of the client population, the size of this cohort is significant and is comparable to the number of clients that would be referred to housing services by the government definitions in Table~\ref{tb.prf}.

\begin{table}[htbp]
  \centering
  \begin{tabular}{l|ccc}
    & Mean & Median & 90th Percentile \\ \hline
    Total Stays & 14.9 & 2.0 & 40.0 \\
    Total Episodes & 1.6 & 1.0 & 3.0 \\
    Tenure (days) & 382.7 & 7.0 & 1450.0 \\
    Usage Percentage & 59.6\% & 93.5\% & 100.0\% \\
    \hline
    \multicolumn{4}{c}{Coverage: 12891/18398 (70.0\%)}\\
    \hline
  \end{tabular}
  \caption{Clients not identified by RAPID}
  \label{tb.pdTmp}
\end{table}

\section{Concluding Remarks}
\label{sec:concl}

Identifying the cohort of people most in need of housing services as quickly as possible is important.  The primary contribution of this paper is to examine how best to do this based on a client's shelter access pattern.  The stays saved and tenure reduction metrics we develop are a new approach for evaluating how well a test for chronic/episodic shelter access satisfies the tradeoff between program efficiency and rapid identification of candidates.

Using these metrics, we present the RAPID tests that we believe are best for identifying chronic/episodic shelter use.  Our tests have the potential to identify clients for housing support almost 200 days sooner than the next fastest definition for chronic homelessness.  The demographic results we present demonstrate that, while fast, RAPID also selects the high impact group of clients that are most in need of housing support.

In the process of evaluating our tests, we have also illustrated the dramatic potential of Housing First services.  Providing a housing intervention at the point in time when a client is identified as a chronic/episodic shelter user has the potential to save hundreds of shelter stays and, in some cases, over 1000 days of shelter interaction {\em per referral}.

While this study has been performed in a Canadian context, our results can be generalized to other jurisdictions.  As noted in Section~\ref{sec:cluster}, variation does exist between homeless populations in different cities.  Ideally, before RAPID is used in a jurisdiction, its performance would ideally be evaluated on a historical shelter dataset from that area.  However, this is not a requirement.  The primary difference between RAPID and other chronic definitions is the reduced time window and stay/episode threshold.  While we believe that our values will work for most cities, program administrators can always adjust RAPID for their circumstances.  For example, using RAPID in areas where even transitional users have a large number of stays, like Philadelphia and Massachusetts \parencite{culhane-dp-2007}, may result in too many housing referrals.  In these cases, administrators could use RAPID as a starting point and increase the threshold and time window parameters over time in order to manage program load.

Finally, we argue that it is worth taking a deeper look at the transitional client population who are not flagged as chronic or episodic.  While their number of stays and episodes are low on average, our analysis demonstrates that there is still a large percentage of the client population that interact with emergency shelters ``under the radar'' for several years.  Any individual who interacts with an emergency shelter even on a very intermittent basis over several years is not in a stable situation.  To provide assistance to this group, a good first step would be to flag clients on their very first return visit to shelter after a long absence for referral to a counselor.

\section{Acknowledgments}
\label{sec:ack}

The authors would like to acknowledge the support of the Natural Sciences and Engineering Research Council of Canada (NSERC), the Calgary Drop-In Centre and the Government of Alberta.  This study is based in part on data provided by Alberta Seniors, Community and Social Services. The interpretation and conclusions contained herein are those of the researchers and do not necessarily represent the views of the Government of Alberta. Neither the Government of Alberta nor Alberta Seniors, Community and Social Services express any opinion in relation to this study.

\printbibliography

\end{document}